\documentclass{PoS}
\usepackage{nico}

\title{Non-perturbative renormalization of 
kaon four-quark operators 
with $n_f=2+1$ Domain Wall fermions}

\ShortTitle{}

\author{
\hfill
\it Edinburgh 2011/05}

\author{Peter Boyle, \speaker{Nicolas Garron} for the RBC-UKQCD collaborations.\\
School of Physics and Astronomy, University of Edinburgh, Edinburgh EH9 3JZ, U.K.\\
        E-mail: \email{nicolas.garron@ed.ac.uk} }

\abstract{We present our strategy and some preliminary results 
for the renormalization of four-quark operators relevant for kaon physics. 
We follow the non-perturbative Rome-Southampton method,
with both exceptional and non-exceptional kinematics. 
We also implement momentum sources and twisted boundary conditions. 
We use an (almost) unitary setup: Domain-Wall valence on 
$n_f=2+1$ Domain-Wall sea and Iwasaki gauge action, 
at two values of the lattice spacing corresponding 
to approximately $0.086$ fm and $0.114$ fm . 
The chiral properties of these fermions play a crucial role 
in this compuation and are studied in detail in this work.
}

\FullConference{The XXVIII International Symposium on Lattice Field Theory, Lattice2010\\
		June 14-19, 2010\\
		Villasimius, Italy}

\begin{document}

\section{Introduction}

Kaon physics has been extensively studied though lattice simulation 
for more than thirty years, and 
thanks to the recent algorithms and hardware developments 
one can now achieve the precision required to constraint 
the standard model and hopefully reveal the effect of new physics.
Recently, $B_K$, the quantity which parametrizes neutral
kaon mixing in the standard model 
has been computed with an accuracy of a few percents~\cite{Aoki:2010pe}.
Nevertheless, other kaon matrix elements are still poorly determined 
although they can have a great impact in the search for new physics or imply strong constraints 
on beyond the standard model (BSM) theories. 
For example, the non-perturbative contributions to neutral kaon mixing 
beyond the standard model have been computed only in the quenched approximation
\cite{Donini:1999nn,Babich:2006bh},
although computation with dynamical fermions are currently underway
and some preliminary studies have been already presented
\cite{Wennekers:2008sg,:2010wq}.
Probably even more importantly, a complete computation of $K\to \pi\pi$ decays with 
dynamical quarks is still missing.
Because the experimental parameters of CP violations are  
very well measured 
($|\epsilon|=(2.228\pm0.011)\times 10^{-3}$ and 
${\rm Re}(\epsilon'/\epsilon)=(1.65\pm0.26)\times 10^{-3}$
~\cite{Nakamura:2010zzi}),
a precise and realistic computation of the relevant matrix elements 
would provide important constraints on the CKM matrix. 
One of the difficulties 
for the lattice implementation comes from the two-body final state.
In the past this problem was usually circumvent by invoking 
the soft pion theorem to relate the two-pion state
to a one-pion state.
But, as it has been shown in~\cite{Li:2008kc},
this approach is not reliable for a precise computation,
mainly because of the poor convergence of chiral perturbation
theory at masses around those of the kaon.
A very important step forward has been made when it 
was realized how the energy shift of a two-particle state
can be computed on the lattice~\cite{Lellouch:2000pv}.
The RBC-UKQCD collaborations have started the computation 
of this decay along this line, and the first results for 
the matrix element of the $\Delta I=3/2$ operators have been presented 
at this conference~\cite{Goode:2011kb}.
An alternative method has been recently presented in~\cite{Laiho:2010ir}.
In this proceeding we present our strategy and some preliminary results
for the renormalization of some of the relevant four-quark operators. 
It has become traditional to use a non-perturbative renormalization scheme
like the Schr\"odinger functional or the RI-MOM scheme. 
Here we use a modified version of the latter, 
following what was done recently for $B_K$~\cite{Aoki:2010pe},
but generalized to other four-quark operators.
In the next section 
we explain what are the operators we consider in this work.
In the third section, we give more details about the numerical techniques,
and preliminary results are presented in the fourth section.

\section{General Framework}
\subsection{Kaon decay}
\begin{figure}[!t]
\begin{center}
\includegraphics[width=4cm]{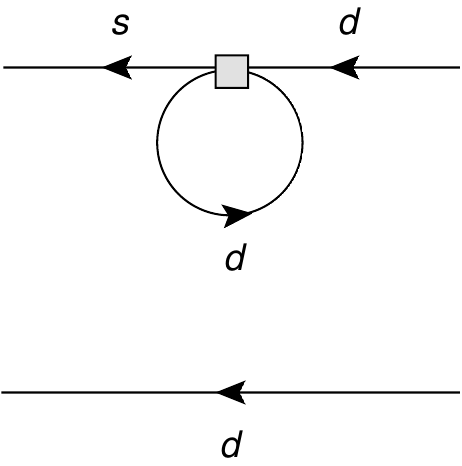}
\caption[]{Example of eye diagram which contribute to $K\to\pi\pi$ decay in the $\Delta I=1/2$ channel.} 
\label{fig:eyediag}
\end{center}
\end{figure}

In the standard model at an energy scale below the charm quark mass,
the dominant non perturbative contributions to  
the effective $\Delta s=1,\; \Delta d=-1$ Hamiltonian can be described
by a linear combination of ten four-quark operators:
two current-current, 
four QCD penguins, 
and four electroweak penguins. 
Among these 10 operators, only seven are actually independent and 
it is useful to classify them according to their chiral and ispospin properties.
Hence one notices that 
they fall into three different representations of $SU(3)_{\rm L}\times SU(3)_{\rm R}$ 
which are $(27,1)$, $(8,1)$ and $(8,8)$, and they can contribute to two ispospin channels 
$\Delta I=3/2$ and $\Delta I=1/2$
(see for example~\cite{Bernard:1989nb,Blum:2001xb}).
We will use the seven-operator renormalization basis defined in~\cite{Blum:2001xb},
in which the operators have the following properties:
\be
\label{eq:Q'basis}
\begin{array}{c|c|c}
\mbox{Operator} & SU(3)_{\rm L}\times SU(3)_{\rm R}& \Delta I \\
\hline
Q'_1     & (27,1)                           & 1/2,\; 3/2 \\
Q'_2,Q'_3,Q'_5,Q'_6  & (8,1)                & 1/2  \\
Q'_7,Q'_8  & (8,8)                          & 1/2, \; 3/2 
\end{array}
\ee
If chiral symmetry was exact, 
the operators of different chirality would not mix under renormalization,
and in the basis described above 
the renormalization matrix would take the block diagonal form:
\be
Z^{\Delta s=1}= \left(
\begin{array}{ccccccc}
Z^{\Delta s=1}_{11} &        &        &         &         &        &         \\
       & Z^{\Delta s=1}_{22} & Z^{\Delta s=1}_{23} & Z^{\Delta s=1}_{25}  & Z^{\Delta s=1}_{26}  &        &         \\
       & Z^{\Delta s=1}_{32} & Z^{\Delta s=1}_{33} & Z^{\Delta s=1}_{35}  & Z^{\Delta s=1}_{36}  &        &         \\
       & Z^{\Delta s=1}_{52} & Z^{\Delta s=1}_{53} & Z^{\Delta s=1}_{55}  & Z^{\Delta s=1}_{56}  &        &         \\
       & Z^{\Delta s=1}_{62} & Z^{\Delta s=1}_{63} & Z^{\Delta s=1}_{65}  & Z^{\Delta s=1}_{66}  &        &         \\
       &        &        &         &         & Z^{\Delta s=1}_{77} &  Z^{\Delta s=1}_{78} \\
       &        &        &         &         & Z^{\Delta s=1}_{87} &  Z^{\Delta s=1}_{88} \\
\end{array}
\right)
\;.
\ee
Furthermore, since ispospin is an exact symmetry in the chiral limit,
for the (27,1) and (8,8) operators, 
it is enough to consider only the $\Delta I=3/2$ parts.
This is numerically advantageous because the  
``eye diagram'' 
(see fig.~\ref{fig:eyediag}) 
which are difficult to compute
can only contribute to $\Delta I=1/2$ processes.

\section{Neutral kaon mixing}
\begin{figure}[!t]
\begin{center}
\includegraphics[width=6cm]{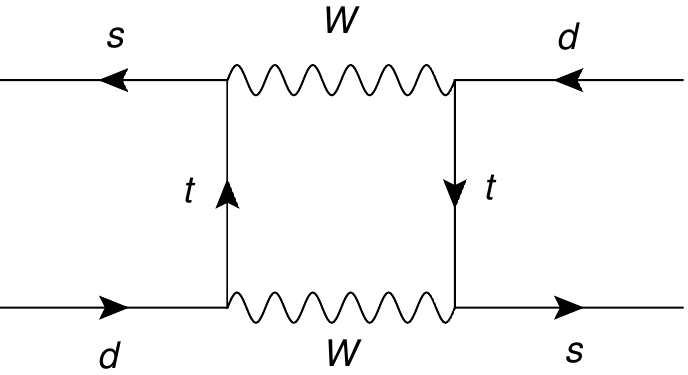}
\caption[]{Example of box diagram contributing to $K-\overline K$ mixing in the Standard model.} 
\label{fig:box}
\end{center}
\end{figure}

In the standard model, neutral kaon mixing is dominated 
by box diagrams like the one shown in figure~\ref{fig:box}.
The non-perturbative
contributions are given by 
$\langle \overline K^0 | O^{\Delta s=2}_{\rm VV+AA} | K^0 \rangle $, 
where $O^{\Delta s=2}_{\rm VV+AA}$ is the parity conserving part 
of $(\overline s \gamma_\mu^L d)(\overline s \gamma_\mu^L d)$.
Beyond the standard model, other operators 
contribute and they are usually given in the so-called SUSY basis
\bea 
\label{eqO1}
O_1^{\Delta s=2}&=&
(\overline s_\alpha \gamma_\mu (1-\gamma_5) d_\alpha)\,
(\overline s_\beta  \gamma_\mu (1-\gamma_5) d_\beta)\,,\\
\label{eqO2}
O_2^{\Delta s=2}&=&
(\overline s_\alpha (1-\gamma_5) d_\alpha)\,
(\overline s_\beta  (1-\gamma_5) d_\beta)\,,\\
\label{eqO3}
O_3^{\Delta s=2}&=&
(\overline s_\alpha  (1-\gamma_5) d_\beta)\,
(\overline s_\beta   (1-\gamma_5) d_\alpha)\,,\\
\label{eqO4}
O_4^{\Delta s=2}&=&
(\overline s_\alpha  (1-\gamma_5) d_\alpha)\,
(\overline s_\beta   (1+\gamma_5) d_\beta)\,,\\
\label{eqO5}
O_5^{\Delta s=2}&=&
(\overline s_\alpha  (1-\gamma_5) d_\beta)\,
(\overline s_\beta   (1+\gamma_5) d_\alpha)\,.
\eea
In this basis $O_1^{\Delta s=2}$ is the standard model operator 
and $O_i^{\Delta s=2}, i>1$ are the BSM ones.
In the SU(3) flavor limit, 
It is straightforward to 
relate $O_4^{\Delta s=2}$ and $O_5^{\Delta s=2}$ to the $\Delta I=3/2$ components 
of the electroweak penguins $Q'_7$ and $Q'_8$, 
which transform under $(8,8)$.
As one can find out from their flavor structures, the remaining operators 
$O_2^{\Delta s=2}$ and $O_3^{\Delta s=2}$ transform under 
$(6,\overline 6)$
~\footnote{
In the literature, we sometimes find the notation $\rm VLL$ for $O_1$, 
$\rm SLL$ for the set $(O_2,O_3)$ and $\rm LR$ for the set $(O_4,O_5)$.}.
Thus, if chiral symmetry is respected, $O_1^{\Delta s=2}$ renormalizes multiplicatively, 
$O_2^{\Delta s=2}$ and $O_3^{\Delta s=2}$ mix together, and so do $O_4^{\Delta s=2}$ and 
$O_5^{\Delta s=2}$.
To simplify the numerical implementation we work in the following basis:
\bea 
\label{eqQ1}
Q_1^{\Delta s=2}&=&(\overline s_\alpha \gamma_\mu d_\alpha)(\overline s_\beta \gamma_\mu d_\beta)+
(\overline s_\alpha \gamma_\mu\gamma_5 d_\alpha)(\overline s_\beta \gamma_\mu \gamma_5 d_\beta) \,,\\
\label{eqQ2}
Q_2^{\Delta s=2}&=&(\overline s_\alpha \gamma_\mu d_\alpha)(\overline s_\beta \gamma_\mu d_\beta)-
(\overline s_\alpha \gamma_\mu\gamma_5 d_\alpha)(\overline s_\beta \gamma_\mu \gamma_5 d_\beta) \,,\\
\label{eqQ3}
Q_3^{\Delta s=2}&=&(\overline s_\alpha d_\alpha)(\overline s_\beta d_\beta)
- (\overline s_\alpha \gamma_5 d_\alpha)(\overline s_\beta \gamma_5 d_\beta) \,,\\
\label{eqQ4}
Q_4^{\Delta s=2}&=&(\overline s_\alpha d_\alpha)(\overline s_\beta d_\beta)
+ (\overline s_\alpha \gamma_5 d_\alpha)(\overline s_\beta \gamma_5 d_\beta) \,,\\
\label{eqQ5}
Q_5^{\Delta s=2}&=&(\overline s_\alpha \sigma_{\mu\nu} d_\beta)(\overline s_\alpha \sigma_{\mu\nu} d_\beta)\,,
\qquad \sigma_{\mu \nu} = {1\over2}[\gamma_\mu,\gamma_\nu] \;.
\eea
The parity conserving part (denoted by a superscript ``+'' )
of the operators (\ref{eqO1})-(\ref{eqO5}) 
can be written in terms of the operators (\ref{eqQ1})-(\ref{eqQ5})
\be
\nonumber
(27,1) \quad {\left[O_1^{\Delta s=2}\right]}^+ = Q_1^{\Delta s=2}
\nonumber
\ee
\begin{minipage}[c]{7.7cm}
\begin{equation}
(6,\overline 6)
\left\{
\begin{array}{rcl}
{\left[O_2^{\Delta s=2}\right]}^+ &=& Q_4^{\Delta s=2} \nonumber\\
{\left[O_3^{\Delta s=2}\right]}^+ &=& -{1\over 2} ( Q_4^{\Delta s=2} - Q_5^{\Delta s=2})\nonumber
\end{array}
\right.
\nonumber
\end{equation}
\nonumber
\end{minipage}
\nonumber
\hspace{1.cm}
\begin{minipage}[c]{6cm}
\hspace{1.cm}
\begin{equation}
(8,8)
\left\{
\begin{array}{rcl}
{\left[O_4^{\Delta s=2}\right]}^+ &=& Q_3^{\Delta s=2} \\
{\left[O_5^{\Delta s=2}\right]}^+ &=& -{1\over 2} Q_2^{\Delta s=2}
\end{array}
\right.
\end{equation}
\end{minipage}
\vspace{0.5cm}\\
It follows from the above considerations that 
$Q_1^{\Delta s=2}$ renormalizes multiplicatively,
$Q_2^{\Delta s=2}$ mixes with $Q_3^{\Delta s=2}$ 
and $Q_4^{\Delta s=2}$ mixes with $Q_5^{\Delta s=2}$. 
We denote the renormalization factors computed in this basis 
by $Z_{ij}^{\Delta s=2}$. 
Moreover, in the $SU(3)$ flavor limit they are some relations between 
the renormalization factors
of the $\Delta s=2$ operators~(\ref{eqQ1})-(\ref{eqQ5}) and those of the 
$\Delta s=1$ operators ~(\ref{eq:Q'basis}). 
For example $O_1^{\Delta s=2}$ and $Q'_1$ have the same 
renormalization factor, and the two by two renormalization matrix
of $(Q_7,Q_8)$ is related to the one of $(Q^{\Delta s=2}_2,Q^{\Delta s=2}_3)$
in the following way
\begin{equation}
\begin{array}{cccccc}
Z^{\Delta s=1}_{77}&=&Z_{22}^{\Delta s=2}   \qquad & Z^{\Delta s=1}_{87}&=&-{1\over2}Z_{32}^{\Delta s=2} \\
Z^{\Delta s=1}_{87}&=&-2 Z_{32}^{\Delta s=2} \qquad & Z^{\Delta s=1}_{33}&=&Z_{88}^{\Delta s=2} \\ 
\end{array}
\end{equation}
\section{Numerical implementation and preliminary results}

The numerical setup of this computation is the same as the one presented 
in details in two recent publications~\cite{Aoki:2010pe,Aoki:2010dy}. 
We use $n_f=2+1$ flavors of domain wall fermion 
on a Iwasaki gauge action at two values of the lattice spacing
$a\sim0.086$ fm and $a\sim 0.114$ fm , corresponding to the lattice volumes
$32\times64\times16$ and $24\times64\times16$, respectively.
On each ensemble the strange sea quark mass is fixed, while
several values of the light sea quark masses have been considered 
(the corresponding unitary pion mass varies in the range $290-420$ MeV).
In this work we consider only the light valence quark masses 
which have the same values as their corresponding sea quarks, 
and perform the chiral extrapolations
linearly in the quark mass. 
This computation was done with 20 configurations, and 100 bootstrap samples.
In addition to the standard RI-MOM scheme, 
we implement also a
scheme with non-exceptional (and symmetric) kinematic,
which exhibits a better infrared behavior~\footnote{
The non-exceptional scheme implemented here is called $\rm SMOM(\gamma_\mu,\gamma_\mu)$
in~\cite{Aoki:2010pe}}
~\cite{Aoki:2007xm,Sturm:2009kb}.
We also employ momentum sources~\cite{Gockeler:1998ye} in order to obtain 
small statistical errors despite the expensive cost of the quark discretization. 
Furthermore, we use twisted periodic boundary conditions, 
which allows us to change smoothly the magnitude of the momentum
without changing its direction (and thus control the $O(4)$ discretization effects)
~\cite{Arthur:2010ht}.
This setup has been used recently used for the computation of $Z_{B_K}$~\cite{Aoki:2010pe}.
Here we generalize this computation to the operators relevant
for neutral kaon mixing beyond the standard model, 
and to the $\Delta I=3/2$ part of $K\to\pi\pi$ decay.
As described in~\cite{Martinelli:1994ty}, 
the Z matrix is essentially the invert of $\Lambda_{ij}=P_j\{O_i\}$,
where $O_i$ is a four-quark operator which belong to the basis given in eqs~(\ref{eqQ1})-(\ref{eqQ5}).
$P_j$ projects onto the Dirac and color structure of the operator $O_j$ 
(and a given flavor structure which depends on the choice of external states).

\begin{center}
\begin{figure}[!t]
\begin{tabular}{cc}
\hspace{-1.5cm}
\includegraphics*[width=9.5cm]{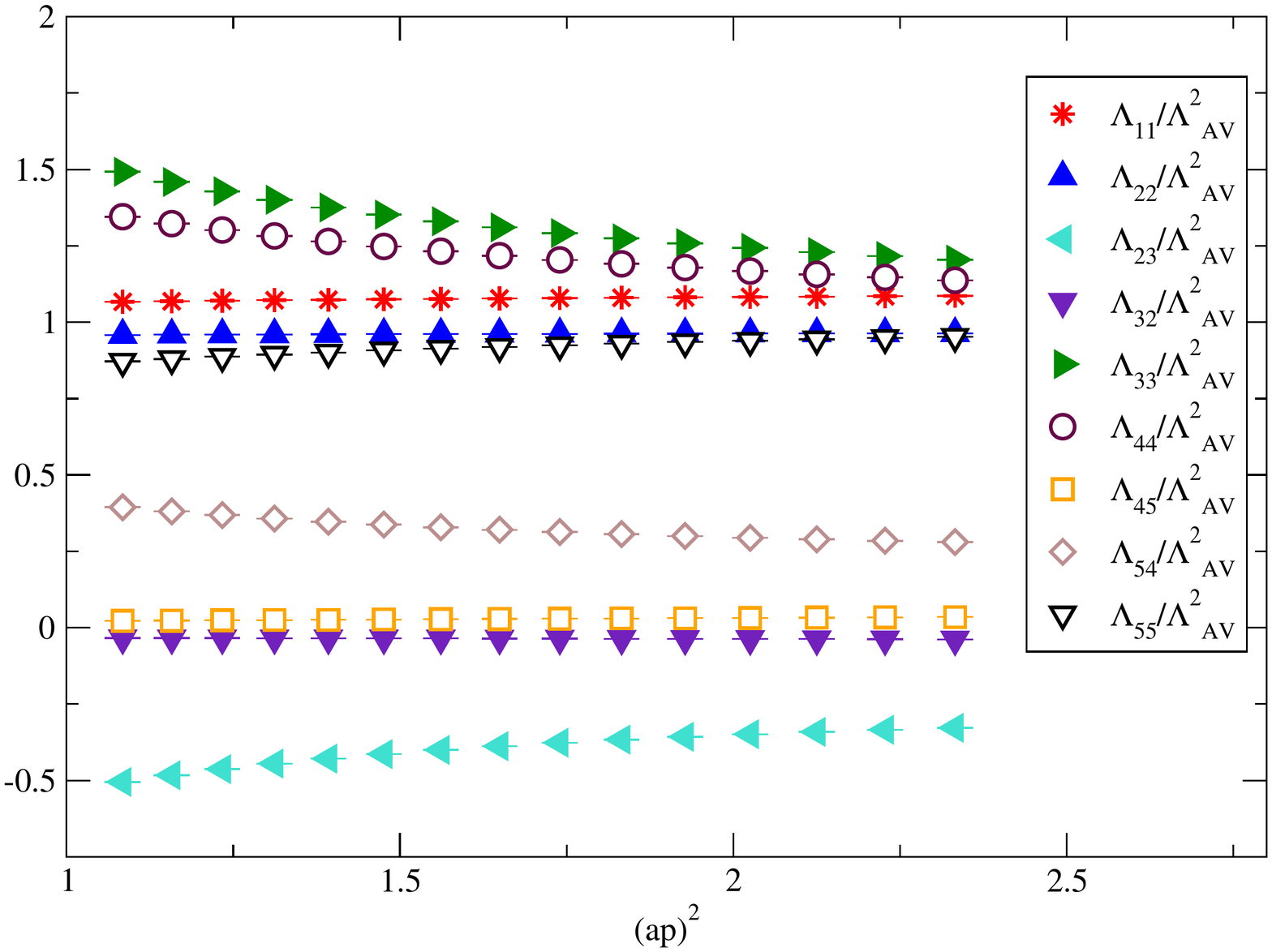} &
\hspace{-1.5cm}
\includegraphics*[width=9.5cm]{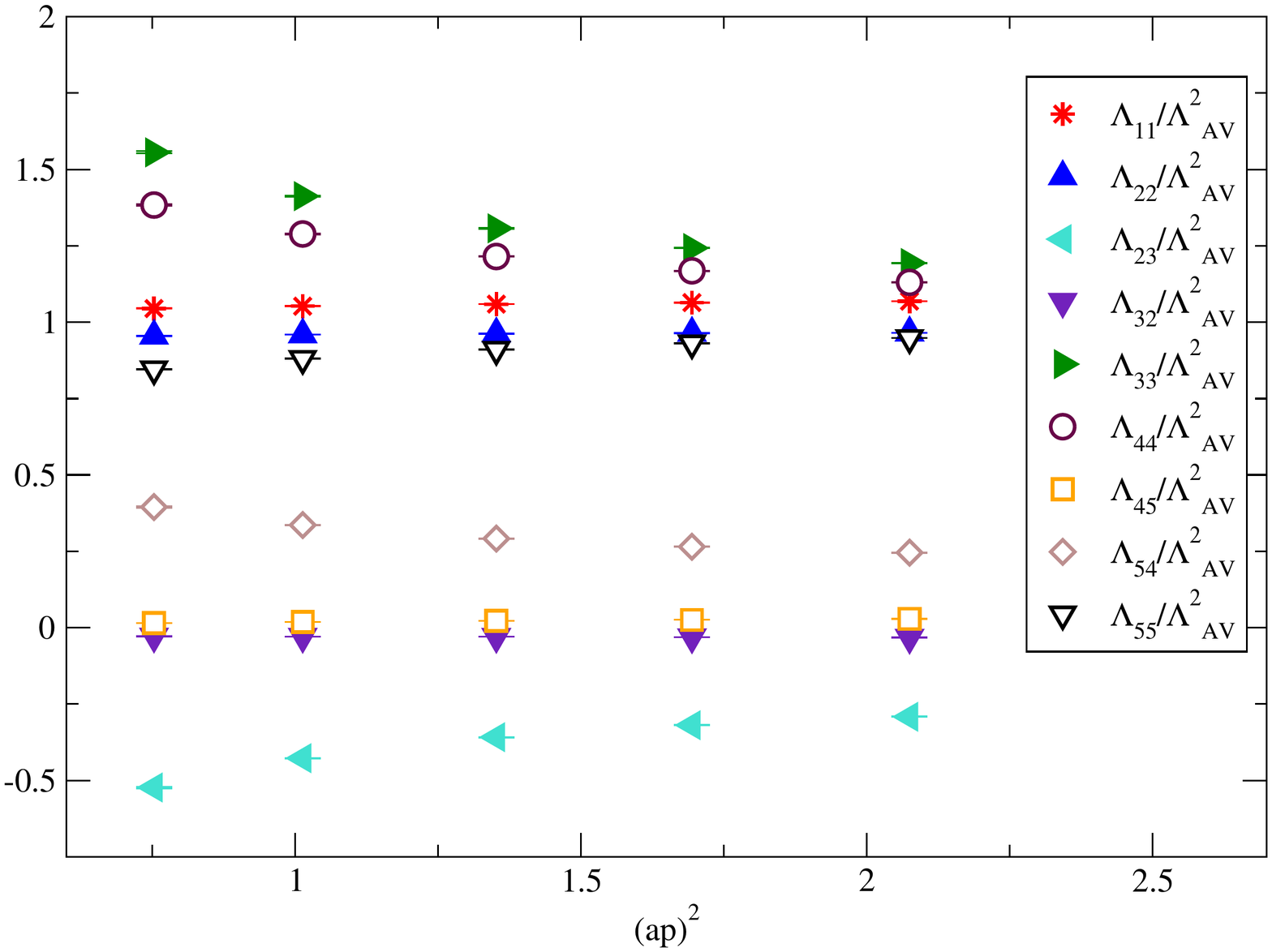}
\end{tabular}
\caption[]{
Physical Green vertex functions for the non exceptional kinematic (SMOM scheme), in the chiral limit.  
On the left panel we show the results for the $24^3$ lattice ($a\sim0.114$ fm)
and on the right for the $32^3$ lattice ($a\sim0.086$ fm).
The error bars are smaller than the symbols.}
\label{figZnonzero}
\end{figure}
\end{center}

\begin{center}
\begin{figure}[!t]
\begin{tabular}{cc}
\hspace{-1.5cm}
\includegraphics*[width=9.5cm]{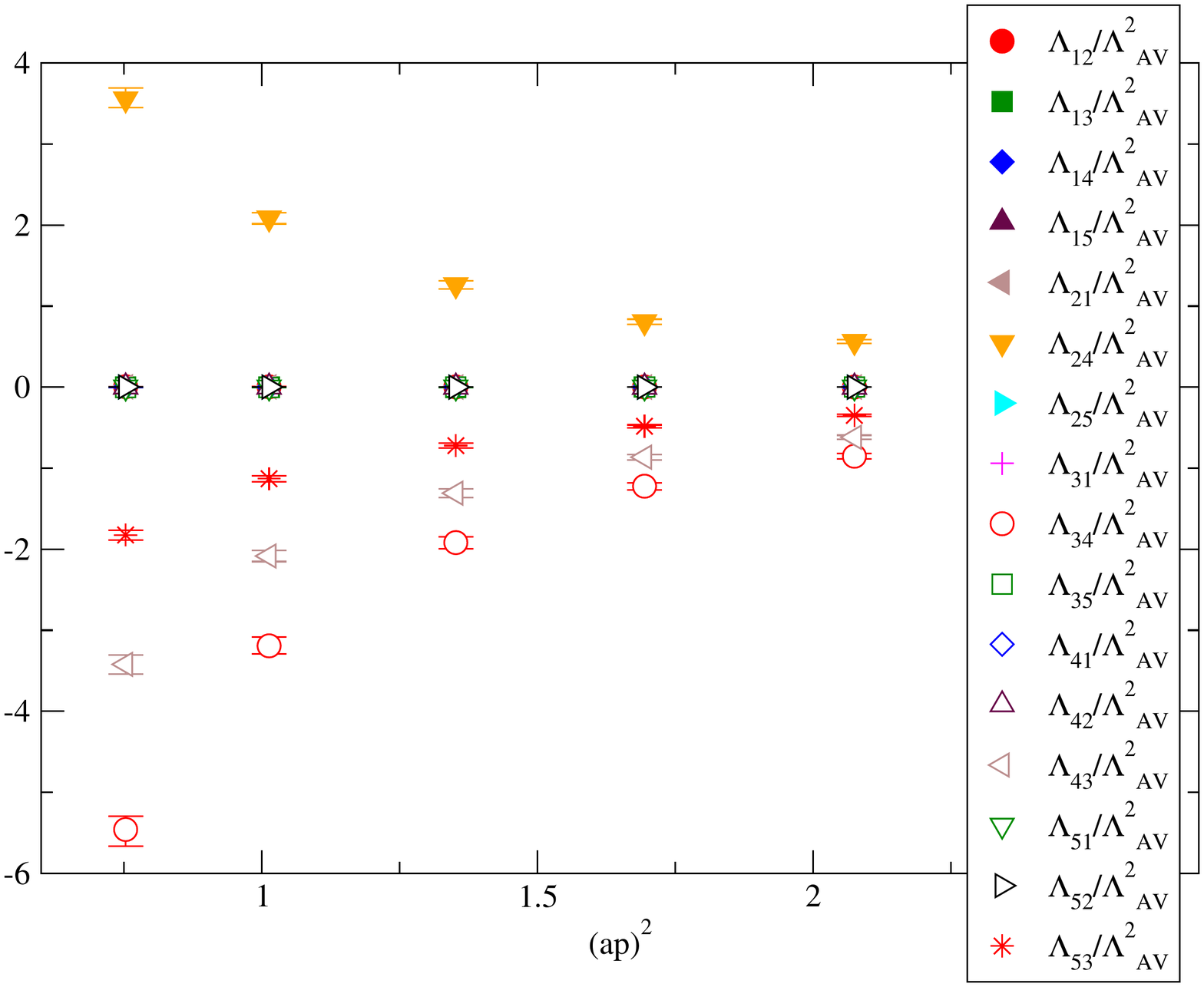} &
\hspace{-1.5cm}
\includegraphics*[width=9.5cm]{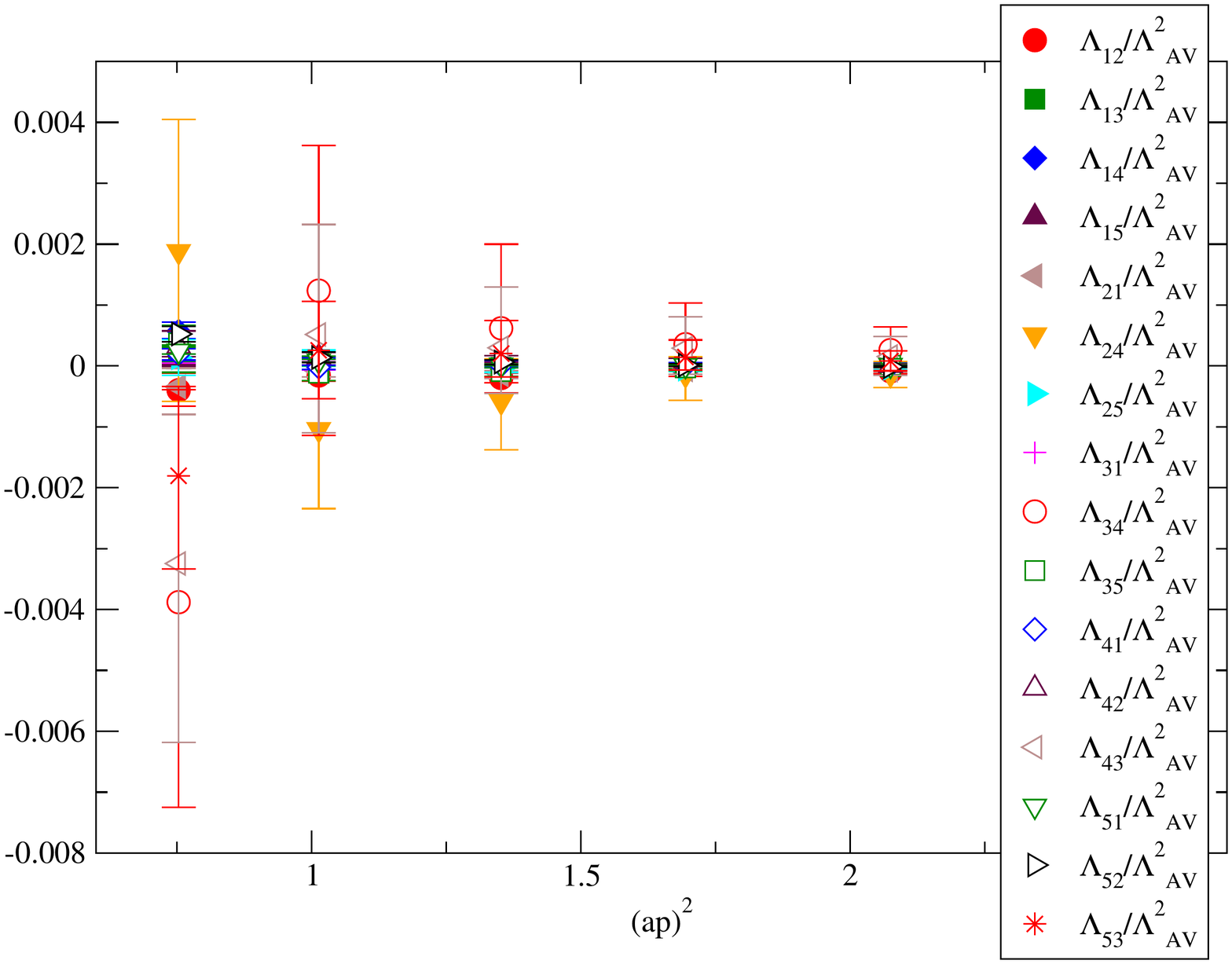}
\end{tabular}
\caption[]{
Chiraly disallowed Green vertex functions on the finest lattice for the exceptional
(left) and non-exceptional (right) kinematic.
We see that for the exceptional kinematic 
some of these matrix elements are suppressed only at high momenta,
while in the non exceptional case they are all zero within the statistical precision.}
\label{figZzero}
\end{figure}
\end{center}

In figures~\ref{figZnonzero} and ~\ref{figZzero}
we show the 
normalized  
Green vertex function $\Lambda_{ij}$
extrapolated to the chiral limit.
The range of simulated momenta corresponds 
to $1.98 \ {\rm GeV} \le p \le 3.29 \ {\rm GeV}$
on the finest lattice and to 
$ 1.80 \ {\rm GeV} \le p \le 2.64 \ {\rm GeV}$ 
on the coarser one.
As expected the use of the momentum sources give us access 
to a very high statistical precision: at a given momentum 
the statistical error is below the permille ($\sim 10^{-4}$ for $Z_{B_K}$).
Thanks to the twisted boundary conditions, the Z factors 
are smooth functions of the momentum (no scatter coming 
from the O(4) discretization effects is visible).
Finally the effect of the non-exceptional kinematic is 
clearly visible in figure~\ref{figZzero}.
The matrix elements shown in these plots should be zero 
if chiral symmetry was exact. 
With the Domain-Wall fermions, this is true only in 
the limit $L_s \to \infty$, so in practice one has 
to check whether the effects of chiral symmetry breaking 
can be seen within the numerical precision.
For the non-perturbative renormalization, this can be complicated 
by the presence of some Goldstone poles, which can affect 
the vertex functions. These poles are suppressed by the use of 
a non-exceptional kinematic. We confirm here an effect already 
seen in ~\cite{Wennekers:2008sg}, that the good 
properties of the Domain Wall action can be obscured by 
a poor choice of kinematic.

\section{Conclusions and outlook}
By combining momentum source with twisted boundary conditions and 
non exceptional kinematic we can obtain the renormalization factors 
of kaon four-quark operators with a very good handle on the different 
kinds or errors: 
the statistical errors are tiny (below the permille),
most of the unwanted infrared effects are suppressed 
and 
the usual scatter coming from the $O(4)$ discretization errors is absent.
Even with this precision, when a non-exceptional kinematic 
is implemented, the non-physical mixing
of the four-quark operators is compatible with zero, thanks
to the good chiral properties of the Domain Wall action.
We are currently extending our computation 
to a larger physical volume discussed in \cite{Goode:2011kb}, where 
the simulated pion mass is significantly smaller (down to 180 MeV). 
We plan to use the step scaling method
introduced in\cite{Arthur:2010ht} in order to enlarge the Rome-Southampton window.
We have also started a computation of the eye diagrams, with the use of stochastic 
sources. \\

We thank our RBC/UKQCD colleagues for many discussions and contributions to this work.

\bibliography{npr}{}
\bibliographystyle{h-elsevier}
\end{document}